# *Implicit Representation of Structural Constraints in ER-to-Relational Transformation: An Analysis of Cardinality Preservation*

Dhammika Pieris

*Department of Scientific Computing*

*Faculty of Computing*

*University of Sri Jayawardenepura, Sri Lanka*
dhammikapieris@sjp.ac.lk

1. Abstract

This study examines the extent to which structural constraints specified in conceptual schemas are represented after transformation to logical schemas. Focusing on the conceptual-to-logical mapping, an Entity–Relationship (ER) model containing binary relationship types is transformed into a Relational Database Schema (RDS). The analysis is conducted under the classical transformation framework in which the logical schema is defined solely by primary key (PK) and foreign key (FK) constraints. Using generalised ER models with variable structural constraint values, the resulting RDS structures are evaluated to determine whether minimum and maximum participation constraints are represented unambiguously. The findings show that, for one-to-one and one-to-many relationships, RDSs do not unambiguously capture minimum participation constraints and do not encode exact maximum participation beyond limited cases. For many-to-many relationships, the schema indicates only that maximum cardinalities exceed one, without preserving exact values. These results clarify the representational limits of standard ER-to-relational transformations and have implications for schema design and constraint enforcement.

## 2. Introduction

When a database development work is undertaken, a conceptual model is drawn to capture and model user requirements of the application domain. The entity-relationship (ER) model (Chen, 1976; Elmasri *et al.*, 2016) that provides a rich set of graphical constructs is widely used for creating conceptual models in the database development process.

An ER model created is then transformed to a relational database schema (RDS) of the relational model (Codd 1970) to design the database. However, much of the information represented on an ER model is often lost during the transformation process (Cuadra *et al.*, 2012; Krishna, 2006) and not frequently reflected on the RDS. For instance, structural constraints (Cuadra *et al.*, 2012; Jumaily *et al.*, 2004), role names, composite attributes, subtype/supertype hierarchies (Goldstein *et al.*, 1999), and some relationship types ) (Pieris *et al.*, 2020) are commonly lost during the transformation process. Therefore, an RDS produced often becomes inconsistent with its source ER model in terms of information representation.

## 3. The Scope

The current study examines the extent to which the structural constraints that are represented on ER models are transformed and represented on the corresponding Relational Database Schemas (RDS). It analyses structural constraint representation under the classical ER-to-relational transformation that uses only primary key and foreign key constraints. The objective is to evaluate what structural constraint information is preserved by the transformation rules themselves, independent of DBMS-specific enforcement mechanisms.

## 4. ER model to Relational model Transformation Method

Various transformation methods are used for transforming an ER schema to an RDS (e.g., Teorey *et al.* (1986), Emalsri and Navathe (1989) as cited in Batini *et al.* (1992), Batini *et al.* (1992), Fahrner *et al.* (1995), Atzeni *et al.* (1999), Dey *et al.* (1999), Goldstein *et al.* (1999), Ramakrishnan *et al.* (2002), Mannino (2007), Ricardo (2004), Ponniah (2007), Coronel *et al.* (2011), and Hoffer *et al.* (2013)). For an updated examination of converting a conventional ER model into the relational model, refer to the work by Grambow in 2023 (Grambow *et al.*, 2023).

For a number of reasons, it is argued that the transformation method proposed by Elmasri *et al.* (2011, pp. 287-299) has been the most successful of the methods so far proposed. It is the most accepted and widely used one.

## 5. Methodology

Relationship types contained in ER models represent structural constraints. A binary relationship type can comprise four structural constraint values placed pairwise on either side of the relationship type. The RDS is analysed to see whether it represents the four structural constraint values unambiguously.

For this purpose, several ER model diagrams are drawn, each containing a binary relationship type: one-to-one, one-to-many or many-to-many, between two regular entity types. After transforming them, the resulting RDS will be critically analysed to extract the four structural constraint values.

Subsequently, the level of representation of the structural constraint values on the RDS will be compared with their actual representation on the corresponding ER model. The values that are not represented and missing on the RDS will be declared.

The analysis commences with a generalised ER model in which the entity types, attributes, relationship types, and structural constraint values are represented in unknown variables.

Consider a binary relationship type $R$ in the following scenario. Assume that $E$ and $S$ are regular entity types with simple attributes, such that Attributes$(E) = \{K_e, A_1, A_2, ...\}$ and Attributes$(S) = \{K_s, A_1, A_2, ...\}$ (Figure 1), where $K_e$ and $K_s$ are the key attributes of the entity types $E$ and $S$, respectively.

Notice that to keep the notation as simple as possible, different symbols are not used for attributes in the two entity types, $E$ and $S$. Instead, similarly notated attributes are used. For instance, attributes of the entity type $E$ are represented as $A_1$, $A_2$,...,. Similarly, the attributes of the entity type $S$ are also represented as $A_1$, $A_2$,...,. However, because they are in different entity types, they are treated as distinct attributes. For instance, the attribute, $A_2$, existing in $E$ should be read as $A_2$ in $E$ or $A_2$ of $E$. While the attribute $A_2$ occurring in $S$ should be read as $A_2$ in $S$. or $A_2$ of $S$. The two attributes, $A_2$ in $E$ and $A_2$ in $S$ are considered two different attributes. The remaining attributes in the two entity types should also be considered as different attributes in the same way as the attribute $A_2$ is deemed.

A structural constraint is denoted by a pair of minimum and maximum values placed in a bracket. Given that the minimum value is $m$, and the maximum value is $x$, the structural constraints between $E$ and $R$ can be given by $(m_1, x_1)$, and the one between $S$ and $R$ by $(m_2, x_2)$, as shown in Figure 1 below.

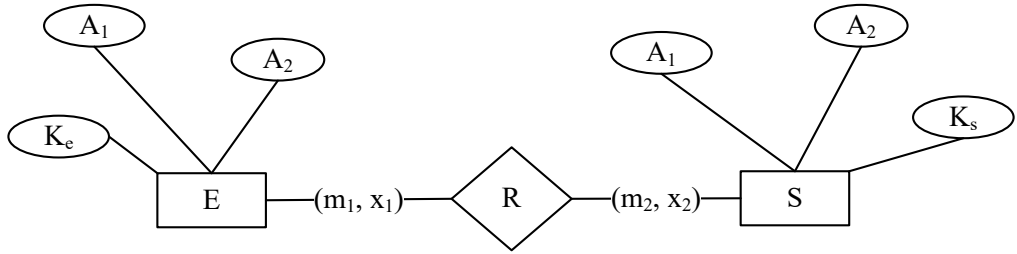

**Figure 1: A binary relationship type, $R$, with structural constraint values in variable form**

The four structural constraint values: $m_1$, $x_1$, $m_2$, and $x_2$, lie in the ranges:

$0 \leq m_1 \leq x_1$ and $x_1 \geq 1$ and

$0 \leq m_2 \leq x_2$ and $x_2 \geq 1$

After applying the transformation algorithm, the transformations of the two entity types are as follows.

$E[K_e, A_1, A_2, \ldots]$

$S[K_s, A_1, A_2, \ldots]$

Accordingly, the relationship type $R$ has not yet been transformed.

Notice that the PKs, $K_e$ and $K_s$, existing in relation schemas are not necessary to be underlined as the character K represents the meaning "Key".

Three main cases are considered for transforming the relationship type $R$. The first, $Case$ 1, assumes that $R$ to be a one-to-one relationship type. The second, $Case$ 2, assumes that $R$ to be a one-to-many relationship type, and the third, $Case$ 3, assumes that $R$ to be a many-to-many relationship type.

### Case 1: R is a one-to-one relationship type

In this case, all the max values of the ER model (Figure 1) must be 1. Accordingly, a new ER model is created with new fixed max values, $x_1 = 1$ and $x_2 = 1$, as shown in Figure 2 below.

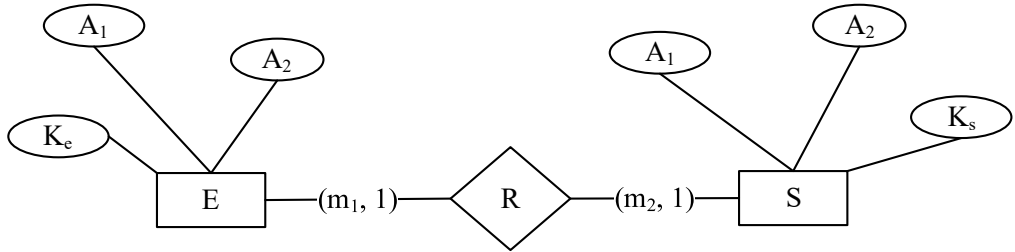

**Figure 2: A binary one-to-one relationship type with participative constraint values in variable form**

However, in this case, the min values: $m_1$ and $m_2$, varies in the ranges: $0 \leq m_1 \leq 1$ and $0 \leq m_2 \leq 1$. In other words, each of $m_1$ and $m_2$ can be either 0 or 1. The possible value pairs for $m_1$ and $m_2$ are shown in the following Table 1.

**Table 1:- Possible values of $m_1$ and $m_2$ in the ER model in Figure 3.8**

| $m_1$ | $m_2$ |
|---|---|
| 1 | 0 |
| 0 | 1 |
| 0 | 0 |
| 1 | 1 |

The RDS of the ER model (Figure 2) will be analysed with min values represented on each row of Table 1. Firstly, the min values: $m_1$ and $m_2$ in Figure 2 are replaced by the fixed values $m_1 = 1$ and $m_2 = 0$ given in the first row of the table. Accordingly, another ER model (Figure 3) is drawn.

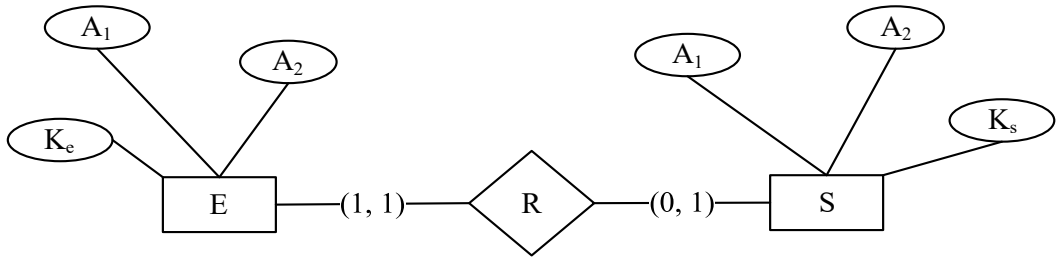

**Figure 3: A binary one-to-one relationship type with the participation constraint value at the entity type $E$'s side being 1**

The entity type $E$ maintains total participation with the entity type $S$ through the relationship type $R$. According to the transformation algorithm, for transforming the relationship type, the PK, $K_s$, of the relation schema, $S$, is included as an FK in the relation schema $E$.

Accordingly, the RDS of the ER model (Figure 3) has become as follows.

$$E[K_e, A_1, A_2, \dots, K_s]$$

$$S[K_s, A_1, A_2, \dots]$$

**Figure 4: The RDS of the ER model in Figure 3 transformed by the transformation algorithm**

The relationship type, $R$, has been represented in the RDS through the FK, $K_s$, included in the relation schema, $E$.

The RDS is analysed to show whether it represents the four structural constraint values: 1, 1, 0, 1, of the ER model (in Figure 3) unambiguously. The analysis is done via four sub-cases, where each case assumes the RDS represents each value unambiguously. Whether the assumption is true or false is shown by analysing the RDS.

First, $Case\ 1 - A$ assumes the RDS represents the max value 1 appearing on the ER model at the entity type $E$'s side from the relationship type $R$. $Case\ 1 - B$ assumes the RDS represents the min value 1 showing at the same side (entity type $E$'s side) of the ER model. $Case\ 1 - C$ assumes the RDS represents the max value 1 appearing on the ER model at the entity type $S$'s side. Finally, $Case\ 1 - D$ assumes the RDS represents the min value 0 appearing on the same side (entity type $S$'s side) on the ER model.

Firstly, for $Case\ 1 - A$, consider the RDS's relation schema, $E$. The PK of the relation schema, $E$, must be non-null and unique across all its tuples. In other words, the PK identifies distinctly the different tuples of the relation schema, $E$.

On the other hand, attributes of the relation schema provide atomic, non-divisible values — i.e., a particular attribute does not provide more than one value for a particular tuple. Thus, tuples are not only distinct but also provide just a single value for each attribute. Consequently, a tuple is associated with only one FK attribute value.

The FK included in the relation schema, $E$, represents a relationship type that $E$ relates with $S$. Accordingly, an FK value occurring in a particular tuple must represent a distinct instance of that relationship type. Thus, a tuple of the relation schema, $E$, associates with only a single instance of the relationship type.

A tuple of the relation schema, $E$, represents an instance of the entity type, $E$. Thus, an instance of the entity type, $E$, associates with a maximum of one instance of the relationship type the FK represents. Consequently, it can be said that the relation schema, $E$, represents the max value 1 that lies at the entity type $E$'s side from the relationship $R$ on the ER model. Thus, the $Case\ 1-A$ assumption is valid. That is, the RDS represents unambiguously the max value 1 that lies at the entity type $E$'s side from the relationship type $R$ on the respective ER model (Figure 3). (This outcome is called the $Case-1-A-Remark$)

Next, for $Case\ 1-B$ consider the relation schema, $E$ again. In the relation schema, $E$, any other attribute other than its PK attribute is not necessarily non-null nor unique in the tuples of the relation schema. Thus, a particular non-PK attribute can be non-null for all the tuples of the relation schema $E$, null for only some tuples of the relation schema, or repeat the same value for some different tuples of the relation schema. Being a non-PK attribute, the FK attribute in the relation schema may behave similarly.

Now, assume that the FK attribute in the relation schema $E$ is null for some tuples of the relation schema, $E$. It means such a tuple belongs to the relation schema $E$ does not associate with any instance of the relationship type the FK represents. In this case, the relation schema $E$ indicates the min value, which lies on the RDS's respective ER model, and at its entity type $E$'s side from the relationship type to be 0. This result nullifies the assumption of $Case\ 1-B$ that the RDS represents the min value 1 showing at the entity type $E$'s side of the ER model. Therefore, it is concluded that the RDS does not represent unambiguously the min value 1 represented on the RDS's respective ER model (Figure 3) at its entity type $E$'s side. (This outcome is called the $Case\ 1-B\ Remark$).

Next, for *Case 1 – C,* consider the RDS in Figure 3 again. Assume a case where each PK value in each tuple of the relation schema, $S$, is included as an FK value in the tuples of the relation schema $E$. However, assume that some of the FK values are repeating in some tuples of the relation schema, $E$.

A PK value of the relation schema $S$ included in the relation schema, $E$, as an FK represents a relationship type that $S$ relates with $E$. Suppose such an FK value repeats in some tuples of the relation schema, $E$. In that case, it indicates that a tuple of the relation schema $S$ associates with multiple instances of the relationship type that FK represents. Assume that a particular repeating FK value provides the maximum number of such repetitions — $n$, ($n > 1$), say.

In this case, the maximum participation of a tuple of the relation schema $S$ in the tuples of the relation schema $E$ via the relationship type is $n$, ($n > 1$). Accordingly, the maximum participation of an instance of the relation schema $S$'s corresponding entity type (the entity type $S$ in the ER model) with instances of the relations schema $E$'s corresponding entity type (the entity type $E$ in the ER model) is $n$, ($n > 1$). It means the max value appearing at the entity type $S$'s side from the relationship $R$ in the ER model is $n$, ($n > 1$), but not 1. This derivation contradicts the assumption *Case 1 – C*.

Therefore, it is concluded that the RDS does not unambiguously represent the actual max value 1 that lies at the entity type $S$'s side of the relationship type $R$ on the respective ER model (Figure 3). (This outcome is called the $Case\ 1 - C - Remark$)

Next, for *Case 1 – D,* consider the RDS in Figure 3 again. When the relation schema $S$'s PK is included in the relation schema, $E$, as an FK, the PK value in each tuple of the relation schema, $S$, can be included in a tuple of the relation schema, *E*. In this case, each tuple of the relation schema, *S*, associates with at least one instance of the relationship type the FK represents.

The relation schema *E* indicates that the minimum participation of the entity type *S* in the relationship type is 1. Thus, in this case, the relation schema *E* indicates the min value that lies on the ER model and at the entity type *S*'s side from the relationship type to be 1. This result contradicts the assumption of the *Case* 1 − *D* that is, the RDS represents the min value 0 appearing at the entity type S's side from the relationship type *R* on the ER model.

Therefore, it is concluded that the RDS does not unambiguously represent the actual min value 0 that lies at the entity type *S*'s side of the relationship type *R* on the respective ER model (Figure 3)(This outcome is called the *Case 1 - D - Remark*).

In conclusion, the RDS in Figure 3 represents only the max value 1 (*Case 1 - A - Remark*) of the structural constraint that lies at the entity type *E*'s side from the relationship type *R* of the RDS's respective ER model (Figure 3). However, it does not represent any of the other three structural constraint values, such that

- The min value 1 (*Case 1 − B - Remark*) that lies at the entity type *E*'s side of the relationship type *R*,

- The next max value, 1 (*Case 1 − C - Remark*) that lies at the entity type *S*'s side of the relationship type *R*, on the ER model in Figure 3.

- The next min value 0 (*Case 1 − D - Remark*) that lies at the entity type *S*'s side of the relationship type *R* on the ER model in Figure 3.

The above result has been obtained for the min values $m_1 = 1$ and $m_2 = 0$ represented in the first row of Table 1 above.

Consider the ER model in Figure 4 below to repeat the test for the next set of fixed min values: $m_1 = 0$ and $m_2 = 1$ given in the second row of Table 3.2.

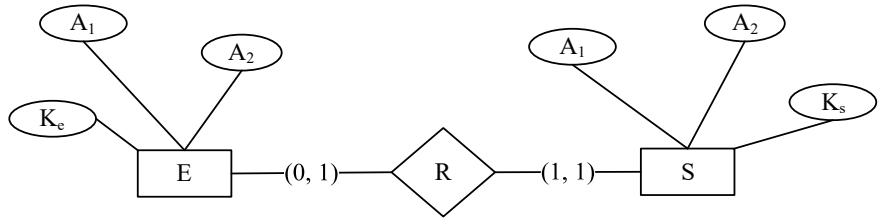

**Figure 5: A binary one-to-one relationship type with the participation constraint value at the entity type *S*'s side being 1**

On the ER model (Figure 5), the entity type $S$ has a total participation with the entity type $E$ through the relationship type $R$. The relationship type $R$ is a one-to-one relationship type. The Step 3 of the existing transformation algorithm transforms it and includes the PK, $K_e$, of the relation schema $E$ as an FK in the relation schema $S$.

Consider the RDS (Figure 6) of the ER model (Figure 5) obtained after the transformation.

$$E[K_e, A_1, A_2, \dots]$$

$$S[K_s, A_1, A_2, \dots, K_e]$$

**Figure 6: The RDS of the ER model in Figure 5 transformed by the existing algorithm**

The relationship type, $R$, has been represented in the RDS by the FK, $K_e$, included in the relation schema, $S$.

It is believed that the RDS in Figure 6 is the same as the one in Figure 4 above in terms of structure. The only difference is the names of the relation schemas, PKs and FKs. For instance, in the previous RDS (the RDS in Figure 4), the name of the relation schema in which an FK is included is $E$, and the name of the included FK is $K_s$. However, in the current RDS, the name of the relation schema in which an FK is included is $S$, and the name of the including FK is $K_e$.

Suppose the same analysis that was undertaken for the previous RDS (Figure 4) is performed for the current RDS (Figure 6). In that case, it is believed that the same conclusions that were drawn for the previous RDS (Figure 4) can be reached for the current RDS (Figure 6) regarding its level of representation of the structural constraints represented on its respective ER model (Figure 5). Accordingly, the following conclusion can be arrived.

The RDS (Figure 6) represents only the max value 1 of the structural constraint that lies at the entity type $S$'s side of the relationship type $R$ on the ER model (Figure 5). However, it does not represent the other three structural constraint values, such that

- The min value 1 that lies at the entity type $S$'s side of the relationship type $R$,

- The next max value, 1, that lies at the entity type $E$'s side of the relationship type $R$, on the ER model in Figure 5.

- The next min value 0 that lies at the entity type $E$'s side of the relationship type $R$.

In this study, it is realised that the same outcome could be reached if this analysis is extend for the other two versions of the ER model in Figure 2. The versions are obtained replacing the values in the 3$^{rd}$ and 4$^{th}$ rows of Table 1 with the ER model's structural constraints' min values: $m_1$ and $m_2$.

Therefore, in summary, suppose in an RDS, a relation schema, $E$, contains another relation schema $S$'s PK as an FK to represent a binary one-to-one relationship type $R$ that exists on the RDS's respective ER model (e.g., Figure 2) between the corresponding entity types, $E$ and $S$. In that case, the relation schema $E$ represents only the $max$ value 1 (e.g., $x_1 = 1$ in Figure 2) lies at the entity type $E$'s side from the relationship type $R$ on the ER model. It does not represent any of the other three values: that is, the two min values (e.g., $m_1, m_2$ in Figure 2) and the remaining max value (e.g., $x_2 = 1$ in Figure 2) represented on the ER model, regardless of the values that $m_1, m_2$ might take in their respective ranges (as shown in Table 1).

### Case 2: R is a one-to-many relationship type

In this case, one of the max values of the ER model in Figure 1 must be 1; for instance, $x_1 = 1$, while the remaining max value must be greater than 1; for instance, $x_2 > 1$, as shown in the ER model Figure 7.

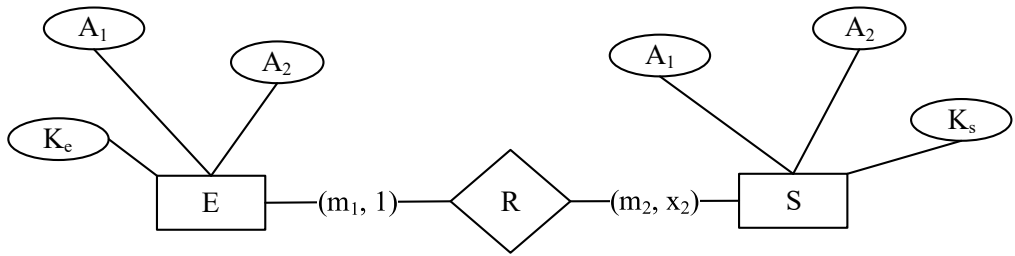

**Figure 7: A binary one-to-many relationship type $R$ that exists between two regular entity types: $E$ and $S$.**

The remaining structural constraint values: $m_1, m_2$, lie in the ranges:

$0 \leq m_1 \leq 1$ and

$0 \leq m_2 \leq x_2$ and $x_2 > 1$

The possible values of $m_1$ and $m_2$ are shown in the following Table 2.

**Table 2:- Possible values that $m_1$ and $m_2$ take, in the ER model in Figure 7**

| $m_1$ | $m_2$ |
|---|---|
| 1 | 0 |
| 0 | 1 |
| 0 | 0 |
| 1 | 1 |
| 1 | n, n > 1 |
| 0 | n, n > 1 |

Table 2 represents six pairs of min values (also called participation constraint values) for the structural constraints of the relationship type $R$ in the ER model (Figure 7).

The RDS of the ER model (Figure 7) will be analysed with min values represented on each row of Table 2. First, another ER model (Figure 8, below) is drawn, replacing the min values: $m_1$ and $m_2$ in the ER model in Figure 7 by the fixed values given in the first row of Table 2.

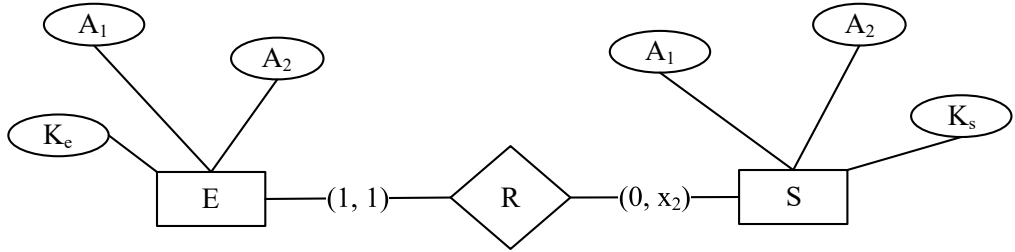

**Figure 8: A binary one-to-many relationship type $R$ that exists between two regular entity types: $E$ and $S$.**

The entity type $E$ maintains total participation with the entity type $S$ through the relationship type $R$. After transformation, the PK, $K_s$, of the relation schema, $S$, is included as an FK in the relation schema $E$.

Consequentely, the RDS of the ER model (Figure 8) can be given as follows.

$$E[K_e, A_1, A_2, \ldots, K_s]$$

$$S[K_s, A_1, A_2, \ldots]$$

**Figure 9: The RDS of the ER model in Figure 8**

The relationship type, $R$, has been represented in the RDS by the FK, $K_s$, included in the relation schema, $E$.

It is believed that the RDS in Figure 9 is the same as the one in Figure 4 created by transforming the ER model in Figure 3, which contained a one-to-one relationship type between two entity types. The relation schemas, PKs, and FKs in the RDS in Figure 9 are the same as those in the RDS in Figure 4 in terms of names and positions in the RDS.

Suppose the same analysis which was already undertaken for the previous RDS (Figure 4) is performed for the current RDS (Figure 9). In that case, the same conclusions can be reached that were drawn for the previous RDS (Figure 4) for the current RDS (Figure 9) regarding its level of representation of the structural constraints represented on its respective ER model (Figure 8).

Accordingly, the following conclusion can be arrived. The RDS (Figure 9) represents only the max value 1 of the structural constraint near the entity type $E$ from the relationship type $R$ on the RDS's respective ER model (Figure 8). However, it does not represent the other three structural constraint values, that is

- The min value 1 that lies near the entity type $E$ from the relationship type $R$,
- The next max value $x_2 > 1$ the one that lies near the entity type $S$ from the relationship type $R$, on the ER model in Figure 8, and
- The next min value 0 that lies near the entity type $S$ from the relationship type $R$.

In this study, it is realised that the same outcome could be reached if this RDSs analysis is extended for the other five versions of the ER model in Figure 7 drawn, replacing the values in the 3rd, 4th, 5th, 6th, and 7th rows of the Table 2 for the ER model's structural constraints' min values: $m_1$ and $m_2$.

In summary, suppose in an RDS, a relation schema, $E$, contains another relation schema $S$'s PK as an FK to represent a binary one-to-many relationship type $R$ that exists on an ER model (e.g., Figure 8). The relationship type $R$ exists between the corresponding entity types, $E$ and $S$. In that case, the relation schema $E$ represents only the $max$ value 1 (e.g., $x_1 = 1$ in Figure 8) that lies at the entity type $E$'s side from the relationship type $R$ on the ER model. It does not represent any of the other three values: that is, the two min values (e.g., $m_1, m_2$ in Figure 8) and the remaining max value (e.g., $x_2 > 1$ in Figure 8) represented on the ER model, regardless of the values that $m_1, m_2$ might take in their respective ranges (as shown in Table 2).

### Case 3: $R$ is a many-to-many relationship type

Consider the many-to-many relationship type, $R$, in the ER model shown in Figure 10. In a binary many-to-many relationship type, both max values are greater than 1 ($x_1 = n$, $(n > 1)$ and $x_2 = q$, $(q > 1)$).

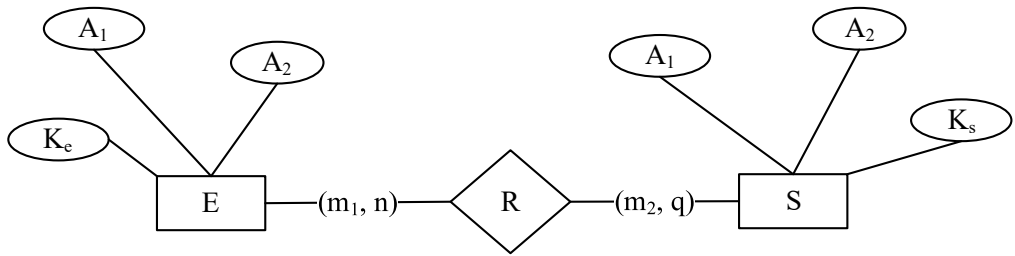

**Figure 10: A binary many-to-many relationship type $R$ that exists between two regular entity types: $E$ and $S$**

To transform the relationship type, $R$, a new relation schema is created by the name of $R$ following the transformation algorithm. The PKs, $K_e$, and $K_s$, of the relation schemas, $E$ and $S$, are included as FKs in $R$. Combining the two key attributes act as the PK of the new relation schema $R$.

$R[K_e, K_s]$

Thus, the complete RDS of the ER model (Figure 10) is as follows.

$$E[K_e, A_1, A_2, ...]$$

$$S[K_s, A_1, A_2, ...]$$

$$R[K_e, K_s]$$

The relation schema, $R$, remains the same for all values of $n$ and $q$ in the ER model. Thus, the relation schema, $R$, only indicates that the two $max$ values of the relationship type, $R$, must be greater than one, that is, $(n > 1)$ and $(q > 1)$.

It does not indicate the exact values of the $max$ cardinality values (i.e., the values of $n$ and $q$).

The relation schema, $R$, also remains the same for all $min$ cardinality values —for instance, $m_1 = 0$ and $m_2 = 0$, or $m_1 = 0$ and $m_2 = 1$, or $m_1 = 1$ and $m_2 = 0$, or $m_1 = 1$ and $m_2 = 1$. Thus, the relation schema R does not represent the $min$ cardinality values.

## 6. Results and Conclusion

Level of representation on the RDSs of the structural constraints of the relationship types, one–to–one, one-to-many, and many–to–many, that are represented on the corresponding ER models are summarised below:

(i). One-to-one relationship type:

Only one of the four structural constraint values, a max value 1, is represented. None of the other three values is represented. For instance, the max value 1 (e.g., $x_1 = 1$, in Figure 3.8), that exists at the side of the entity type, in which relation schema that a PK is included as an FK, is represented by the same relation schema in the RDS. However, none of the other three values (e.g., $m_1$, $m_2$, and $x_2 = 1$, in Figure 2) are represented.

(ii). One-to-many relationship type:

Similar to a one-to-one relationship type, in this case also, only one of the four structural constraint values, a max value 1, is represented. None of the other three values is represented. For instance, the max value 1, $(x_1 = 1)$ (e.g., Figure 7), that exists at the side of the entity type, and in which relation schema that a PK is included as an FK, is represented by the same relation schema in the RDS. However, none of the other three values (e.g., $m_1$, $m_2$, and $x_2$ in Figure 7) are represented.

(iii).   Many-to-many relationship type:

The fact that the two max values are greater than one is represented in the corresponding RDS by the new relation schema itself. Nonetheless, the exact values of the max values are not represented. Furthermore, none of the min values (e.g., $m_1$ and $m_2$ in Figure 10) are represented.

The above analysis is the first time it could be able to figure out precisely what structural constraint values, which are represented on an ER model in association with a binary relationship type, could be transformed and represented on the RDS.

A method needs to be found on how the values that the current method does not transform can be represented on the RDS. However, the existing format and the rules of the RDS should not be compromised in finding a way. If it could be found, the next task is to modify the existing algorithm to cater to the new method.

## 7. References

Atzeni, P*., et al.* (1999). *Database Systems: Concepts, Languages and Architectures* London: Mc Graw Hill.

Batini, C*., et al.* (1992). *Conceptual database design: an Entity-relationship approach*: Benjamin-Cummings Publishing Co., Inc. Redwood City, CA, USA ©1992.

Chen, P. P. S. (1976). The entity-relationship model: toward a unified view of data. *ACM Trans. Database Syst, 1*(1), 9-36. doi:10.1145/320434.320440

Coronel, C*., et al.* (2011). *Database systems: Design, implementation, and management*: Course Technology Cengage Learning.

Cuadra, D*., et al.* (2012). Guidelines for representing complex cardinality constraints in binary and ternary relationships. *Software and Systems Modeling*, 1-19. doi:10.1007/s10270-012-0234-3

Dey, D*., et al.* (1999). Improving database design through the analysis of relationships. *ACM Transactions on Database Systems (TODS), 24*(4), 453-486.

Elmasri, R*., et al.* (2011). *Fundamentals of Database Systems*. New York: Addison Wesley.

Elmasri, R*., et al.* (2016). *Fundamentals of Database Systems* (7 ed.). Delhi: Pearson.

Fahrner, C*., et al.* (1995). A survey of database design transformations based on the Entity-Relationship model. *Data & Knowledge Engineering, 15*(3), 213-250.

Goldstein, R. C*., et al.* (1999). Data abstractions: Why and how? *Data & Knowledge Engineering, 29*(3), 293-311.

Grambow, G*., et al.* (2023). *A Practical Automated Transformation of Entity Relationship Models to Relational Models.* Paper presented at the Fifteenth International Conference on Advances in Databases, Knowledge, and Data Applications, Barcelona, Spain.

Hoffer, J. A*., et al.* (2013). *Modern Database Management* (11 ed.). Boston USA: Pearson

Jumaily*, et al.* (2004). Applying a Fuzzy approach to relaxing cardinality constraints. In F. Galindo, Takizawa, M., Traunmüller, R (Ed.), *Database and Expert Systems Applications* (Vol. 3180, pp. 654-662): Springer, Berlin, Heidelberg.

Krishna, M. (2006). Retaining semantics in relational databases by mapping them to rdf. *Proceedings of the Web Intelligence and Intelligent Agent Technology Workshops (WI-IAT)*, 303-306. doi:10.1109/WI-IATW.2006.114

Mannino, M. V. (2007). *Database design, application development, and administration*: McGraw-Hill New York.

Pieris, D*., et al.* (2020). ER Model Partitioning: Towards Trustworthy Automated Systems Development. *International Journal of Advanced Computer Science and Applications(IJACSA), 11*(6). doi:10.14569/IJACSA.2020.0110638

Ponniah, P. (2007). *Data modeling fundamentals: a practical guide for IT professionals*: Wiley-Blackwell.

Ramakrishnan, R*., et al.* (2002). *Database Management Systems* (2 ed.). Boston, MA: Mc Graw Hill.

Ricardo, C. M. (2004). *Database Illuminated* (1 ed.). London: Jones and Bartlett.

Teorey, T. J*., et al.* (1986). A logical design methodology for relational databases using the extended entity-relationship model. *ACM Comput. Surv., 18*(2), 197-222. doi:10.1145/7474.7475